\documentclass[conference]{IEEEtran}

\IEEEoverridecommandlockouts

\usepackage{amsmath}
\usepackage{amssymb}
\usepackage{mathrsfs}

\usepackage{balance}   

\usepackage{cite} 

\usepackage{ifpdf}  

\usepackage{multicol}
\usepackage{makeidx}
\usepackage{color}
\usepackage{pseudocode}
\usepackage{epsf}
\usepackage{subfigure}

\ifCLASSINFOpdf
   \usepackage[pdftex]{graphicx}
   \graphicspath{{../fig/pdf/}}
   \DeclareGraphicsExtensions{{.pdf}}
\else
   \usepackage[dvips]{graphicx}
   \graphicspath{{../fig/eps/}}
   \DeclareGraphicsExtensions{{.eps}}
\fi

\makeatletter
\def\normalsize{\@setfontsize{\normalsize}{10bp}{10.00pt}}
\normalsize
\makeatother

\def\refname{\normalsize \centerline{\bf REFERENCES} }

\DeclareMathAlphabet{\mathpzc}{OT1}{pzc}{m}{it}

\newcommand{\be}{\begin{eqnarray}}
\newcommand{\ee}{\end{eqnarray}}

\begin{document}

\title{Noncoherent Analog Network Coding \\ using LDPC-coded FSK \vspace{-0.15cm} }

\author{
Terry Ferrett and
Matthew C. Valenti,\\
West Virginia University, Morgantown, WV, USA.
\vspace{0cm}
}

\maketitle

\begin{abstract}
Analog network coding (ANC) is a throughput increasing technique for the two-way relay channel (TWRC)
whereby two end nodes transmit simultaneously to a relay at the same time and band, 
followed by the relay broadcasting the received sum of signals to the end nodes.
Coherent reception under ANC is challenging due to requiring oscillator synchronization
for all nodes, a problem further exacerbated by Doppler shift.
This work develops a noncoherent M-ary frequency-shift keyed (FSK) demodulator implementing ANC.
The demodulator produces soft outputs suitable for use with capacity-approaching channel codes and
 supports information feedback from the channel decoder.
A unique aspect of the formulation is the presence of an infinite summation in the received symbol
probability density function.
Detection and channel decoding succeed when the truncated summation contains a sufficient number of terms.
Bit error rate performance is investigated by Monte Carlo simulation, considering modulation orders two, four and eight, channel coded and uncoded operation, and with and without information feedback from decoder to demodulator.
The channel code considered for simulation is the LDPC code defined by the DVB-S2 standard.
To our knowledge this work is the first to develop a noncoherent soft-output demodulator for ANC.
\end{abstract}

\IEEEpeerreviewmaketitle


\section{Introduction}

In the \emph{two-way relay channel} (TWRC) two \emph{end nodes} exchange information
through an intermediate \emph{relay node}.
The end nodes have no direct radio link to each other, and are both in range of the relay.
\emph{Physical-layer network coding} (PNC) \cite{zhang2:2006} is a transmission scheme
which reduces the number of time slots required for information exchange.
The exchange is divided into the \emph{multiple-access} (MA) phase and \emph{broadcast} (BC) phase.
In the MA phase, the sources transmit simultaneously, and the relay receives the electromagnetic
sum of transmissions.
In the BC phase, the relay broadcasts the combination of signals to the
end nodes, each of which detect the information transmitted by the opposite end node.

A primary distinction between PNC schemes is the forwarding technique applied
by the relay \cite{zhang:2008}.
In the case that the relay amplifies and forwards the signal received from the end nodes
during the MA phase, the forwarding technique is termed \emph{PNC over an infinite field} or
\emph{analog network coding} (ANC)\cite{katti:2007}.
When the relay demodulates and optionally performs channel decoding, the forwarding technique is
referred to as \emph{PNC over a finite field} or \emph{digital network coding} (DNC) \cite{ferrett:2010},
as the relay detects and forwards information symbols over a discrete and finite set, 
such as an M-ary frequency-shift keyed (FSK) constellation.

A significant challenge in developing practical PNC receivers for the TWRC
is achieving \emph{phase synchronization} between the three nodes in the network, 
which is required for coherent reception.
Variations in transmitted signal frequencies due to oscillator imperfections
and Doppler shifts make synchronization challenging.
While it may be straightforward to synchronize oscillators between two nodes,
the third will still exhibit an offset that must be taken into account in
receiver design.
Phase synchronization challenges motivate the investigation of noncoherent reception.

Our previous work developed a soft-output noncoherent M-FSK demodulator for DNC at the relay in the TWRC
\cite{valenti:2009}  \cite{vtf:2011} \cite{ferrett:2013}.
The current work develops a \emph{soft-output noncoherent M-FSK demodulator for ANC at the end nodes in the TWRC},
the first of its kind to our knowledge.
The demodulator supports power-of-two modulation orders and produces log-likelihood ratios (LLRs)
suitable for use with capacity approaching soft-decision decoding techniques.
The performance of LDPC channel coding coupled with ANC is investigated in this work.
As a noncoherent formulation, the demodulator is capable of operating without
any knowledge of the channel and without phase synchronization between the end node
and relay oscillators.

Previous work on ANC analyzes achievable transmission rates, compares
with other TWRC protocols, and develops noncoherent receivers.
An analysis of the achievable rates for ANC for a variety of network topologies is considered in \cite{maric:2010}.
Closed form expressions for the bit-error rate of noncoherent FSK in the passive RFID channel are
derived in \cite{he:2011}.  
The passive RFID channel is analytically similar to the broadcast channel under ANC, as both
consider a signal transmitted over two Rayleigh fading channels, an instance of \emph{double Rayleigh fading} \cite{lu:2011}.
The relationship between bit error rate, transmission rate, and transmit power for the ANC TWRC is analyzed in
\cite{yang:2016}, forming the basis for a rate and power adaptation scheme.
A noncoherent receiver for the ANC TWRC using uncoded differential PSK modulation is developed in \cite{song:2010}.
The BER of the receiver is derived and an optimal power allocation scheme is developed assuming
constant fading coefficients per frame. 

The following organization is applied for the rest of the work.
Section II describes the system model.
Section III presents the ANC demodulator, developing the probability distribution of the symbols and bits
received at the end nodes.
Section IV provides the simulation procedure and performance results used to investigate the performance
of the developed demodulator.
Concluding remarks are provided in Section V.

\section{System Model}\label{sec:sysm}

\begin{figure}[t]
\centering
\vspace{-0.5cm}
\includegraphics[width=9.3cm]{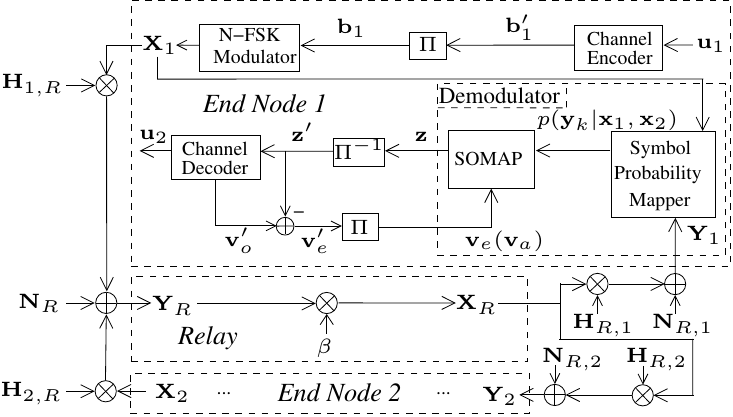}
\caption{System Model - Analog Network Coded Two-way Relay Channel.  The configuration of End Node 2 is identical to 1,
and has been omitted from the figure.}
\label{fig:sysm}
\vspace{-3mm}
\end{figure}

This section describes the system model assumed for derivation and simulation of the ANC soft-output end node demodulator.
The channel model is described, followed by end node modulation with and without channel coding.
Relay operation is described.
End node reception using the developed demodulator with and without channel decoding is 
described.
Symbol and frame synchronization is assumed throughout.
The system model is shown in Fig. \ref{fig:sysm}.

\subsection{Transmission by End Nodes}

Two \emph{end nodes} $\mathcal{N}_1$ and $\mathcal{N}_2$ generate information bit sequences $\mathbf{u}_i = [ u_{1,i}, ..., u_{K,i}], \ i \in \{1,2\}$ having length $K$.
Under channel coded operation, each $\mathbf{u}_i$ is encoded by an LDPC code having rate $r_S$, yielding a length $L = K/r_S$ channel codeword, denoted by $\mathbf{b}'_i = [b_{1,i} ..., b_{L,i}]$.
Under uncoded operation, $\mathbf{b}_i = \mathbf{u}_i$ and $L = K$.
The codeword is passed through an interleaver, modeled as a permutation matrix $\mathbf{\Pi}$ having dimensionality $L \times L: \mathbf{b}_i = \mathbf{b}'_i \mathbf{\Pi}$.
The number of bits per symbol is $\mu = \log_2 M$.
The codeword $\mathbf{b}_i$ is partitioned into $N_q = L/\mu$ sets of bits.
Each set of $\mu$ bits is mapped to an $M$-ary FSK symbol $q_{k,i} \in \mathcal{D}$, where $k$ denotes the symbol index, and $i$ denotes the end node, and $\mathcal{D} = \{ 0, ..., M-1 \}$ denotes the set of integer indices corresponding to each FSK tone.

The modulated signal transmitted by each end node during interval $k T_s \leq t < (k+1) T_s $ is
\vspace{-0mm}
\begin{align}\label{eqn:td_sym}
s_{k,i}(t) =  \sqrt{ \frac{2}{T_s} }
\cos
\left[
  2 \pi
  \left(
    f
    + \frac{q_{k,i}}{T_s}
    \right )
    (t - kT_s)
    \right]
\end{align}

\noindent where $f$ is the end node carrier frequency and $T_s$ is the symbol period.
A vector model is assumed where each vector dimension models the output of a matched
filter tuned to a particular FSK frequency, and the frequency spacing is chosen such that 
the tones are orthogonal. 
A transmitted symbol is represented by the column vector $\mathbf{x}_{k,i}$,
where $k$ is the symbol interval and $i$ denotes the $i$-th end node.
The vector $\mathbf{x}_{k,i}$ has length $M$, and contains a $1$ at vector position corresponding to the transmitted
tone $q_{k,i}$, and $0$ elsewhere.
The frame of modulated symbols transmitted by end node $\mathcal{N}_i$ is represented by the matrix of symbols $\mathbf{X}_i = [\mathbf{x}_{1,i},...,\mathbf{x}_{N_q,i}]$, having dimensionality $M \times N_q$.

\subsection{Channel Model for Multiple Access Phase}

In the MA phase, a frequency-flat fading model is assumed where the channel gains are independent for every symbol period.
The gain from node $\mathcal{N}_i$ to the relay during signaling interval $k$ is modeled as a circularly
symmetric complex jointly Gaussian random variable denoted by $h_{k,i,R} \sim \mathcal{N}_c( 0, \mathcal{E}_i )$, where $\mathcal{E}_i$ is the variance.
In polar form the gain is represented as $h_{k,i,R} = \alpha_{k,i,R} e^{j \theta_{k,i,R}}$, where $\alpha_{k,i,R}$ is the Rayleigh distributed amplitude and $\theta_{k,i,R}$ is the phase, uniformly distributed between $\left[0, 2\pi \right)$.
The variances are chosen such that the energy received at the relay from transmission by node $\mathcal{N}_i$ is $\mathcal{E}_i$
\vspace{-0mm}
\begin{eqnarray}\label{eqn:fad_ampl_energy}
E[|h_{k,i,R}|^2] = E[\alpha^2_{k,i,R}] = \mathcal{E}_i .
\end{eqnarray}
The received signal at the relay after transmission of a single symbol frame by each end node is
\vspace{-0mm}
\begin{align} \label{eqn:rec_sym}
\mathbf{Y}_R = \mathbf{X}_1 \mathbf{H}_{1,R} + \mathbf{X}_2 \mathbf{H}_{2,R} + \mathbf{N}_R
\end{align}

\noindent where $\mathbf{H}_{i,R}$ is a square diagonal matrix of fading coefficients with dimensions $N_q \times N_q$
modeling the fading between end node $\mathcal{N}_i$ and the relay.
The matrix takes value $h_{k,i,R}$ at row and column $(k,k)$ and $0$ elsewhere.
The matrix $\mathbf{N}_R$ is an $M \times N_q$ noise matrix.
Denote the $k$-th column of $\mathbf{N}_R$ by $\mathbf{n}_{k,R}$.
Each column is composed of zero-mean circularly symmetric complex jointly Gaussian random variables having covariance matrix $N_0 \mathbf{I}_M$; i.e., $\mathbf{n}_{k,R} \sim \mathcal{N}_c(\mathbf{0}, N_0 \mathbf{I}_M)$.
$N_0$ is the one-sided noise spectral density, and $\mathbf{I}_M$ is the $M$-by-$M$ identity matrix.
A single signaling interval is represented by a single column of $\mathbf{Y}_R$ and is denoted by $\mathbf{y}_{k,R}$.
In terms of this definition, $\mathbf{Y}_R =  [\mathbf{y}_{1,R}, ..., \mathbf{y}_{k,R}, ..., \mathbf{y}_{N_q,R}]$.

\subsection{Analog Network Coding at the Relay}

During the BC phase, the purpose of the relay is to broadcast the frame of received symbols $\mathbf{Y}_R$ to the end nodes after scaling to satisfy the power constraint.
Consider a single received symbol $\mathbf{y}_{k,R}$.
The relay forms a symbol to transmit by scaling $\mathbf{y}_{k,R}$ as 
\vspace{-0mm}
\begin{align}
\mathbf{x}_{k,R} = & \beta \mathbf{y}_{k,R} \nonumber \\
 = & \beta ( h_{k,1,R} \mathbf{x}_{k,1} + h_{k,2,R} \mathbf{x}_{k,2} + \mathbf{n}_{k,R})
\end{align}
\noindent where $\mathbf{x}_{k,R}$ denotes the $k$-th symbol formed for transmission by the relay, and $\beta$ is a real-valued scaling factor which constrains the average transmitted energy. 
The relay forms a frame of symbols to broadcast to the end nodes as 
$\mathbf{X}_R = [\beta \mathbf{y}_{1,R}, ..., \beta \mathbf{y}_{N_q,R}] = [\mathbf{x}_{1,R}, ..., \mathbf{x}_{{N_q}, R}]$.

The value of the scaling factor $\beta$ which normalizes the transmitted energy
depends on the statistics of the received symbols.
Under noncoherent operation, the exact values of the fading coefficients $h_{k,i}$ are not known at the relay.
It is assumed that the relay can estimate the statistics of the fading coefficients and additive noise.
Specifically, the variances of the fading coefficients $\mathcal{E}_i$ and additive noise $N_0$ are assumed known
through estimation. 

Consider reception of a single symbol $\mathbf{y}_{k,R}$ at the relay.
The total energy of the received symbol is 
\vspace{-0mm}
\begin{align}
\mathcal{E}_{k} = \sum_{m=0}^{M-1} |y_m|^2
\end{align}
\noindent where $m$ denotes the $m$-th dimension of $\mathbf{y}_{k,R}$.
The average energy received during a symbol period is computed as
\vspace{-0mm}
\begin{align}
\mathcal{\bar{E}}_R  =& E\left[ \sum_{m=0}^{M-1} |y_m|^2 \right] \nonumber \\
 = & N_0 M + \mathcal{E}_1 + \mathcal{E}_2
\end{align}
\noindent where it is assumed that the end nodes transmit all symbols with equal probability.
The average energy transmitted by the relay is normalized to unity by setting the scaling factor as
\vspace{-0mm}
\begin{align}
\beta = \frac{1}{\sqrt{ N_0 M + \mathcal{E}_1 + \mathcal{E}_2  } }.
\end{align}
\noindent Since the scaling factor depends only on the statistics of the fading coefficients rather than
the exact values, it is constant for a particular realization of the statistics.

\subsection{End Node Reception}

The goal of reception at each end node is to detect the information bits transmitted by the opposite end node.
During the BC phase, each end node receives the symbol frame broadcast by the relay after the frame has traversed a fading channel.
Demodulation and optional channel decoding is performed to detect the desired information bits. 
Each end node knows the symbol frame it transmitted during the multiple access phase, and this information is used to compute the conditional probability of receiving particular symbols from the opposite node.

The frame received at end node $\mathcal{N}_i$ during the broadcast phase is
\vspace{-0mm}
\begin{align} \label{eqn:rec_sym_en}
\mathbf{Y}_i = \mathbf{X}_R \mathbf{H}_{R,i} + \mathbf{N}_i
\end{align}
\noindent where $\mathbf{H}_{R,i}$ denotes the diagonal matrix of fading coefficients for the channel between the relay and end node $\mathcal{N}_i$, having dimensions $N_q \times N_q$, and $\mathbf{N}_i$ is an $M \times N_q$ noise matrix having the same distribution as $\mathbf{N}_R$.
The channel gains forming the diagonal for matrix $\mathbf{H}_{R,i}$ are denoted by $h_{k,R,i}$ and are distributed as circularly symmetric complex jointly Gaussian $\mathcal{N}_c( 0, \mathcal{E}_R )$.
The matrix $\mathbf{H}_{R,i}$ takes value $h_{k,R,i}$ at row and column $(k,k)$ and $0$ elsewhere.

The demodulator takes as input the symbols received from the relay $\mathbf{Y}_R$, the symbols transmitted by the end node during the multiple access phase $\mathbf{X}_i$, and a-priori probability (APP) information regarding the bits under detection $\mathbf{v}_a$.
As output, the demodulator produces a-posteriori information regarding the bits under detection $\mathbf{z}$.
The a-posteriori information is deinterleaved to produce $\mathbf{z}' = \mathbf{z} \mathbf{\Pi}^{-1}$ and passed to the channel decoder.
The decoder refines the estimate of $\mathbf{z}'$, producing a-posteriori information $\mathbf{v}_o'$.  
The decoder input is subtracted from the decoder output to produce extrinsic information $\mathbf{v}_e' = \mathbf{v}_o' - \mathbf{z}'$ which is interleaved to produce $\mathbf{v}_e = \mathbf{v}_e' \mathbf{\Pi}$ and returned to the demodulator.
The decoder output becomes the demodulator a-priori input $\mathbf{v}_a = \mathbf{v}_e$.

The end nodes are assumed to know the average noise power $N_0$ and fading statistics in the form of variances $\mathcal{E}_1$, $\mathcal{E}_2$ and $\mathcal{E}_R$.
This information can be obtained through a variety of techniques such as pilot symbols and control channels between the relay and end nodes.
Knowledge of the noise power and fading statistics are assumed known in the formulation of the end node demodulator.

Formulation of the demodulator is described in Section \ref{sec:demod}.
Details of the channel decoder have been described at length in the literature and will not be discussed in this work.

\section{Noncoherent End Node Demodulator}\label{sec:demod}

This section develops the end node soft-output demodulator.
The probability distribution of the symbols received at the end nodes is developed,
followed by the model for iterative demodulation and decoding at the end node.
Since demodulation is performed on a single symbol at a time, for the purpose of formulating the demodulator,
we may drop the dependence on symbol period $k$ throughout to simplify the notation.

\subsection{End Node Received Symbol Distribution}

Consider a single received symbol at end node $\mathcal{N}_i$
\begin{align}
\mathbf{y}_i =& h_{R,i} \mathbf{x}_R + \mathbf{n}_{i} \nonumber \\
=& \beta h_{R,i} (h_{1,R}\mathbf{x}_1 + h_{2,R} \mathbf{x}_2 + \mathbf{n}_R) + \mathbf{n}_{i}.
\end{align}

\noindent The term $\mathbf{x}_R$ is formed by the sum of three vectors, each having components which are circularly symmetric complex jointly Gaussian random variables, and all components are independent.
Since the sum of complex jointly Gaussian random variables is also complex and jointly Gaussian, the components of 
$\mathbf{x}_R$ are distributed $\mathcal{N}_c(0, \sigma^2_m)$ where $\sigma^2_m$ is the variance of the $m$-th vector component $x_{m,R}$.
The values of the variances depend on the symbols transmitted by the end nodes
\begin{align}
\sigma^2_m = 
\begin{cases}
N_0 & x_{m,1}=0, x_{m,2}=0 \\
N_0 + \mathcal{E}_1 & x_{m,1}=1, x_{m,2}=0 \\
N_0 + \mathcal{E}_2 & x_{m,1}=0, x_{m,2}=0 \\
N_0 + \mathcal{E}_1 + \mathcal{E}_2 & x_{m,1}=1, x_{m,2}=1.
\end{cases}
\end{align}

Now consider the distribution of the product of the fading coefficient $h_{R,i}$ and
the symbol transmitted by the relay
\begin{align}
 \boldsymbol{\mu} &= h_{R,i} \mathbf{x}_R  \nonumber \\
&= [ \ h_{R,i} x_{0,R},\ ...,\ h_{R,i} x_{M-1,R}\ ]^T. \nonumber \\
&= \beta [ \ \alpha_0 e^{i \theta_0},\ ...,\ \alpha_{M-1} e^{i \theta_{M-1}}\ ]^T.
\end{align}
\noindent Each component of $\boldsymbol{\mu}$ is the product of two independent circularly symmetric complex Gaussian random
variables, which yields the \emph{complex double Gaussian distribution} having PDF \cite{moura:2012}
\begin{align}\label{eq:cdg1}
p_{\mu_m}(\alpha_m, \theta_m) = \frac{2 \alpha_m}{\pi \mathcal{E}_R \sigma^2_m} K_0 \left( \frac{2\alpha_m}{\sqrt{\mathcal{E}_R} \sigma_m} \right)
\end{align}
\noindent where $K_0(\cdot)$ is the \emph{modified Bessel function of the second kind} \cite{NIST:DLMF}. 

We now derive the distribution of the received symbol which does not depend on knowledge of the fading amplitudes and
phases.
Denote the amplitudes of the components of $\boldsymbol{\mu}$ as $\boldsymbol{\alpha} = [\alpha_0,\ ...,\ \alpha_{M-1}]$
and the phases as $\boldsymbol{\theta} = [\theta_0,\ ...,\ \theta_{M-1}]$.
The distribution of the received symbol conditioned on $\boldsymbol{\alpha}$ and $\boldsymbol{\theta}$ becomes
\begin{align}
p(\mathbf{y}|\boldsymbol{\alpha}, \boldsymbol{\theta}) =& \left(\frac{1}{\pi N_0}\right)^M \exp \left[ -\frac{||\mathbf{y} - \boldsymbol{\mu}||^2}{N_0} \right] \nonumber \\
=& \left(\frac{1}{\pi N_0}\right)^M \ \prod_{m=0}^{M-1}\exp \left[ -\frac{|y_m - \beta \alpha_m e^{i \theta_m}|^2}{N_0} \right].
\end{align}

Note that the joint distribution of the amplitude and phase  given by (\ref{eq:cdg1}) is the product of marginal distributions
$p_{\mu_m}(\alpha_m, \theta_m) = p(\alpha_m) p(\theta_m)$, where 
\begin{align}
p(\alpha_m) = \frac{4 \alpha_m}{\mathcal{E}_R \sigma^2_m} K_0 \left( \frac{2\alpha_m}{\sqrt{\mathcal{E}_R} \sigma_m} \right) 
\end{align}
and $p(\theta_m) = \frac{1}{2\pi}, \ 0 \leq \theta_m < 2 \pi$, thus, we may marginalize over the amplitude and phase separately.

Marginalizing over the phases yields
\vspace{-0mm}
\begin{align}
&p(\mathbf{y}|\boldsymbol{\alpha}) = \int_0^{2 \pi} p(\mathbf{y}|\boldsymbol{\alpha}, \boldsymbol{\theta}) p(\boldsymbol{\theta}) d \boldsymbol{\theta} \nonumber\\
=& \left(\frac{1}{\pi N_0}\right)^M \ \prod_{m=0}^{M-1} \int_0^{2 \pi} \exp \left[ -\frac{|y_m - \beta \alpha_m e^{i \theta_m}|^2}{N_0} \right] \frac{1}{2 \pi} d\theta_m \nonumber \\
=& \left(\frac{1}{\pi N_0}\right)^M \ \exp \left[ -\frac{\sum\limits_{m=0}^{M-1} |y_m|^2 }{N_0} \right] \times ... \nonumber \\
&\prod_{m=0}^{M-1} \exp \left[- \frac{\beta^2 \alpha_m^2}{N_0} \right] I_0 \left( \frac{2 \beta \alpha_m |y_m| }{N_0} \right)
\end{align}
\noindent where $I_0(\cdot)$ is the \emph{modified Bessel function of the first kind} \cite{NIST:DLMF}.

Marginalizing over the amplitudes yields
\begin{align}\label{eq:a_pdf_1}
&p(\mathbf{y})= \int_0^{\infty} p(\mathbf{y}|\boldsymbol{\alpha}) p(\boldsymbol{\alpha}) d \boldsymbol{\alpha} \nonumber\\
=& \left(\frac{1}{\pi N_0}\right)^M \ \exp \left[ -\frac{\sum\limits_{m=0}^{M-1} |y_m|^2 }{N_0} \right] \left( \frac{4}{\mathcal{E}_R} \right)^M \times ... \nonumber \\
&\prod_{m=0}^{M-1} \frac{1}{\sigma^2_{m}} \int_0^{\infty} \exp \left[- \frac{\beta^2 \alpha_m^2}{N_0} \right] I_0 \left( \frac{2 \beta \alpha_m |y_m| }{N_0} \right) \times ... \nonumber \\
&\alpha_m K_0 \left( \frac{2 \alpha_m}{\sqrt{\mathcal{E}_R} \sigma_{m}} \right) d \alpha_m
\end{align}

\noindent For the purpose of performing the integration given by (\ref{eq:a_pdf_1}), we may neglect the terms outside the integral for a moment, yielding 
\begin{align}\label{eq:a_pdf_int}
\int_0^{\infty} \alpha_m \exp \left[- \frac{\beta^2 \alpha_m^2}{N_0} \right] I_0 \left( \frac{2 \beta \alpha_m |y_m| }{N_0} \right)
\nonumber \times ...\\ 
K_0 \left( \frac{2 \alpha_m}{\sqrt{\mathcal{E}_R} \sigma_{m}} \right) d \alpha_m.
\end{align}

To perform the integration, we represent the modified Bessel function of the first kind as a series \cite{NIST:DLMF}
\hspace{-0mm}
\begin{align}\label{eq:b1_series}
I_0(x) = \sum_{n=0}^\infty \frac{x^{2n}}{4^n n!^2}
\end{align}
\noindent After substituting (\ref{eq:b1_series}) into (\ref{eq:a_pdf_int}), the integral becomes
\begin{align}\label{eq:a_pdf_2}
 \int_0^{\infty} \alpha_m \exp \left[- \frac{\beta^2 \alpha_m^2}{N_0} \right] \sum_{n=0}^\infty  \frac{(c_1 \alpha_m)^{2n}}{4^n n!^2}
  K_0 \left( \frac{2 \alpha_m}{\sqrt{\mathcal{E}_R} \sigma_{m}} \right) d \alpha_m
\end{align}

\noindent where $c_1 = 2 \beta |y_m| / N_0$.  
Factoring out constants with respect to the integration and rearranging, (\ref{eq:a_pdf_2}) becomes
\begin{align}\label{eq:a_pdf_3}
\sum_{n=0}^\infty  \frac{c_1^{2n}}{4^n n!^2} \int_0^{\infty} \alpha_m^{2n+1} \exp \left[- \frac{\beta^2 \alpha_m^2}{N_0} \right] 
  K_0 \left( \frac{2 \alpha_m}{\sqrt{\mathcal{E}_R} \sigma_{m}} \right) d \alpha_m
\end{align}

\noindent Define $c_2 = \beta^2/N_0$.  We then make the change of variable $u=c_2 \alpha_m^2$ and $du = 2 c_2 \alpha_m d \alpha_m$.
Then $\alpha_m = \sqrt{u/c_2}$ and $d \alpha_m = du/(2 c_2 \alpha_m)$.
Substituting the change of variable into (\ref{eq:a_pdf_3})
\begin{align}\label{eq:a_pdf_4}
\frac{1}{2} \sum_{n=0}^\infty  \frac{c_1^{2n}}{4^n n!^2 c_2^{n+1}} \int_0^{\infty} u^n
\exp(-u)  
 K_0 \left( c_4 \sqrt{u} \right)
du
\end{align}
\noindent where $c_4 = 2/(\mathcal{E}_R \sigma^2_m \sqrt{c_2})$.
Applying integration formula (6.643-3) in \cite{grad:2007}, (\ref{eq:a_pdf_4}) becomes
\begin{align}\label{eq:whit_int}
\frac{ \exp \left( c_4^2/8 \right) }{ c_4 }
\sum_{n=0}^\infty \frac{ c_1^{2n} }{ 4^n } W_{-(n+1/2), 0} \left( \frac{c_4^2}{4} \right)
\end{align}
\noindent where $W_{a,b}(x)$ is the Whittaker-W function \cite{NIST:DLMF} having parameters $a$ and $b$ and argument $x$.

Substituting the result of integration (\ref{eq:whit_int}) into (\ref{eq:a_pdf_1}) yields the
PDF of the received symbol having no dependence on the channel state
\begin{align}\label{eq:rec_pdf}
&p(\mathbf{y}) = ... \nonumber \\
&\left( \frac{1}{\pi \sqrt{N_0} \sqrt{\mathcal{E}_R} \beta} \right)^M
\exp \left[ \sum\limits_{m=0}^{M-1} \left( -\frac{|y_m|^2}{N_0} + \frac{N_0}{2 \mathcal{E}_R \sigma^2_{m} \beta^2} \right)\right]
\times \nonumber \\
&\prod_{m=0}^{M-1} \frac{1}{\sigma_{m}} 
\sum\limits_{n=0}^\infty \left( \frac{|y_m|^2}{N_0} \right)^n 
W_{-(n+1/2), 0} \left( \frac{N_0}{\mathcal{E}_R \sigma^2_{m} \beta^2} \right).
\end{align}
\noindent This expression is suitable for performing noncoherent soft output detection at the end nodes.
The PDF contains an infinite summation, which is truncated for implementation.

\subsection{Iterative Demodulation and Decoding}

The end node demodulator maps the symbols received from the relay during the broadcast phase
to log-likelihood ratios of the bits transmitted by the opposite end node.
In the following section, without loss of generality, consider reception at end node $\mathcal{N}_1$,
where the goal is to recover the information sequence $\mathbf{u}_2$ transmitted by $\mathcal{N}_2$.
Iterative decoding is performed whereby the channel decoder feeds information back to the demodulator,
which refines the bit estimates and sends them back to the channel decoder.
A hard decision is made on the bits after the specified iteration count has been reached.

The soft mapper (SOMAP) \cite{benedetto:1998} operates on a symbol-by-symbol basis, transforming symbol probabilities $p(\mathbf{y}|\mathbf{x}_1=\mathbf{a},\mathbf{x}_2)$ to the set of $\mu$ log-likelihood ratios associated with each bit mapped to symbol $\mathbf{x}_2$.
The term $\mathbf{a}$ is the symbol transmitted by the receiving end node during the symbol period under consideration,
which is available, since the end node knows the data that it transmitted.
The SOMAP takes as input the symbol probabilities and a-priori information fed back from the channel decoder about the bits mapped to the symbols $\mathbf{v}_a$.
The SOMAP produces a-posteriori log-likelihood ratios of the bits mapped to the channel symbols $\mathbf{z}$.
On the first iteration, no decoding has been performed, and the bit probabilities are assumed equally likely, yielding
$\mathbf{v}_a = \mathbf{0}$.

The a-priori log-likelihood ratio of the $m$-th bit mapped to input symbol $\mathbf{x}_2$ is
\vspace{-0mm}
\begin{align}\label{eq:input_bit}
v_k = \log \frac{ P(u_k = 1; I) }{ P(u_k = 0; I) }, \ 0 \leq k \leq \mu - 1 .
\end{align}

\noindent The a-posteriori SOMAP output is the log-likelihood ratio of the $k$-th bit mapped to $\mathbf{x}_2$
\vspace{-0mm}
\begin{align} \label{eq:somap_out_llr}
z_k = \log \frac{ P(u_k = 1; O) }{ P(u_k = 0; O) }, \ 0 \leq k \leq \mu - 1 .
\end{align}

\noindent The SOMAP input is transformed to output according to
\vspace{-0mm}
\begin{align}\label{eq:somap_out_distribution_symbolic}
P(u_k=\ell; O) = \sum_{\begin{subarray}  (\mathbf{x}_2: u_k = \ell \end{subarray}}  p(\mathbf{y} |\mathbf{x}_1=\mathbf{a}, \mathbf{x}_2) \prod_{\begin{subarray} jj=0 \\j \neq m \end{subarray}}^{\mu-1} P( u_j ; I)
\end{align}

\noindent Substituting (\ref{eq:input_bit}) into the expression for output (\ref{eq:somap_out_distribution_symbolic}),
\vspace{-0mm}
\begin{align}\label{eq:somap_out_distribution_specific}
P( u_k = \ell; O) =  \sum_{\begin{subarray} (\mathbf{x}_2: u_k = \ell \end{subarray}}  p(\mathbf{y} |\mathbf{x}_1=\mathbf{a}, \mathbf{x}_2)  \prod_{\begin{subarray} jj=0 \\j \neq m \end{subarray}}^{\mu-1}  \frac{ e^{ u_j v_j} }{ 1+e^{v_j} } 
\end{align}

The SOMAP out log-likelihood ratio may be found by combining (\ref{eq:somap_out_distribution_specific}) and (\ref{eq:somap_out_llr}):
\vspace{-0mm}
\begin{align}\label{eq:somap_out_llr_full}
z_k = \log \frac{ \displaystyle\sum_{\begin{subarray} (\mathbf{x}_2: u_k = 1 \end{subarray}} p(\mathbf{y} |\mathbf{x}_1=\mathbf{a}, \mathbf{x}_2)  \prod_{\begin{subarray} jj=0 \\j \neq m \end{subarray}}^{\mu-1}   e^{ u_j  v_j}  }
 { \displaystyle\sum_{\begin{subarray} (\mathbf{x}_2: u_k = 0 \end{subarray}} p(\mathbf{y} |\mathbf{x}_1=\mathbf{a}, \mathbf{x}_2)  \prod_{\begin{subarray} jj=0 \\j \neq m \end{subarray}}^{\mu-1}    e^{u_j v_j}  }
\end{align}
\noindent where the term $(1+e^{v_j})$ cancels in the ratio.
When implementing (\ref{eq:somap_out_llr_full}), simplification using the \emph{max-star} operator
provides numerical stability.
The max-star operator is defined as
\vspace{-0mm}
\begin{align}
\underset{i}{\operatorname{max} \hspace{-0.5mm}*} \{ x_i \} = \log \left\{ \sum_i e^{ x_i } \right\}
\end{align}

\noindent where the binary max-star operator is  $\max*(x,y) = \max(x,y) + \log( 1 + e^{ -|x-y| } ) $ and
multiple arguments are recursive. For example, in the case of three arguments, max-star becomes $\max*(x,y,z) = \max*( x, \max*(y,z) )$.
Applying the max-star operator to (\ref{eq:somap_out_llr_full})
\vspace{-0mm}
\begin{align} \label{eq:somap_out_llr_maxstar}
 z_k & =  \underset{\begin{subarray} (\mathbf{x}_2: u_k = 1 \end{subarray}}{\operatorname{max}  \hspace{-0.5mm} *} \left[ \log p(\mathbf{y} | \mathbf{x}_1=\mathbf{a}, \mathbf{x}_2) + \sum_{\begin{subarray} jj=0 \\ j \neq k\end{subarray}}^{\mu-1} u_j v_j\right] \nonumber \\ & -\underset{\begin{subarray} (\mathbf{x}_2: u_k = 0 \end{subarray}}{\operatorname{max}  \hspace{-0.5mm} *} \left[ \log p(\mathbf{y} | \mathbf{x}_1=\mathbf{a}, \mathbf{x}_2) + \sum_{\begin{subarray} jj=0 \\ j \neq k\end{subarray}}^{\mu-1} u_j v_j  \right].
\end{align}

\noindent A non-iterative demodulator does not use decoder feedback, and is implemented using (\ref{eq:somap_out_llr_maxstar}) setting all $v_j = 0$.

The term $\log p(\mathbf{y} | \mathbf{x}_1=\mathbf{a}, \mathbf{x}_2)$ in (\ref{eq:somap_out_llr_maxstar}) is computed by taking the logarithm of (\ref{eq:rec_pdf}), yielding
\begin{align}\label{eq:rec_pdf_log}
\log& \ p(\mathbf{y}|\mathbf{x}_1, \mathbf{x}_2) = \sum_{m=0}^{M-1} \left[ \frac{N_0}{2 \mathcal{E}_R \sigma^2_m \beta^2} - \log\sigma_m \right] + ...\nonumber \\
&\sum_{m=0}^{M-1}\underset{0 \leq n \leq N_t}{\operatorname{max} \hspace{-0.5mm}*} 
\biggl[ 2n \log|y_m| - n \log N_0 + ... \nonumber \\
&\hspace{22mm}\log W_{-(n+1/2),0} \left( \frac{N_0}{\sigma^2_R \sigma^2_m \beta^2} \right) \biggr].
\end{align}
\noindent where the infinite series has been truncated to a finite number of terms $N_t$.
Note that the following terms in (\ref{eq:rec_pdf})
\begin{align}
\left( \frac{1}{\pi \sqrt{N_0} \sqrt{\mathcal{E}_R} \beta} \right)^M
\exp \left[ \sum\limits_{m=0}^{M-1} \left( -\frac{|y_m|^2}{N_0} \right)\right]
\end{align}
\noindent cancel in the ratio given by (\ref{eq:somap_out_llr_maxstar}), and are not included in (\ref{eq:rec_pdf_log}).
Demodulator performance as a function of the truncation length $N_t$ is investigated in Section \ref{sec:demodperf}.

\section{Demodulator Performance}\label{sec:demodperf}
This section presents Monte Carlo simulated error rate performance for the demodulator derived in Section \ref{sec:demod}.
Error rate performance is simulated using different values of modulation order, demodulator summation terms, with and without channel coding, with and without decoder feedback to the demodulator (BICM vs BICM-ID) and signal-to-noise ratio.
Both end nodes and the relay transmit each each symbol with unit energy, making the variance of the fading coefficients
$\mathcal{E}_1 = \mathcal{E}_2 = \mathcal{E}_R = 1$.
The channel code considered is the LDPC code defined by the \emph{DVB-S2} standard \cite{dvbs2:2013}.

\begin{figure}[!t]
  \centering
  \includegraphics[width=9cm]{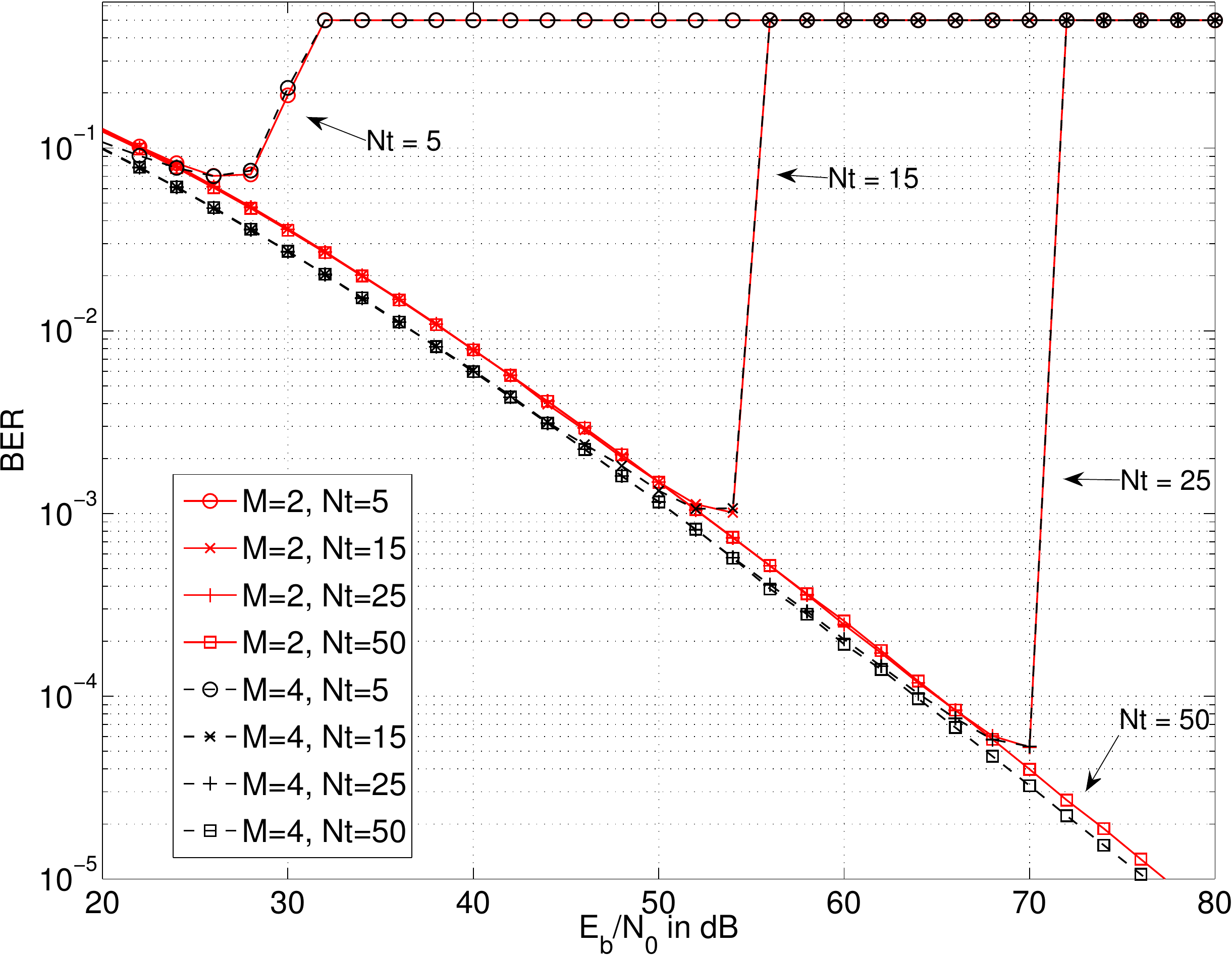}
  \caption{Bit error rate performance with no channel coding at the end node in the two-way relay channel broadcast phase
under Rayleigh fading.
The modulation orders considered are $M=\{2,4\}$.
The number of demodulator infinite series terms considered are $N_t=\{5,15,25,50\}$.}
  \label{fig:uncoded}
\vspace{-3mm}
\end{figure}

\subsection{Error Rate Performance}

The results of error rate simulation are presented in this subsection.
All uncoded simulations use frame size $K=2048$ bits.
Coded simulations use the DVB-S2 LDPC code with codeword length $L=16200$ and rate
$K/L=1/2$.
All coded simulations apply $100$ decoding iterations.
When no information is fed back from the decoder to the demodulator (BICM),
all decoding iterations are performed by the decoder.
When information is fed back from the decoder to the demodulator (BICM-ID),
a single channel decoder iteration is performed for every iteration between the 
decoder and the demodulator.
BICM-ID is performed for modulation orders $M>2$, as there it provides no benefit for
$M=2$.
The FSK modulation orders considered are $M=\{2,4,8\}$.
Computation of the infinite series in the expression for received symbol probabilities (\ref{eq:rec_pdf_log})
is truncated to finite values $N_t = \{ 5, 10, 15, 20, 50\}$.
For all simulations, enough trials are run to yield smooth error rate curves.

 \begin{figure}[!t]
   \centering
   \includegraphics[width=9cm]{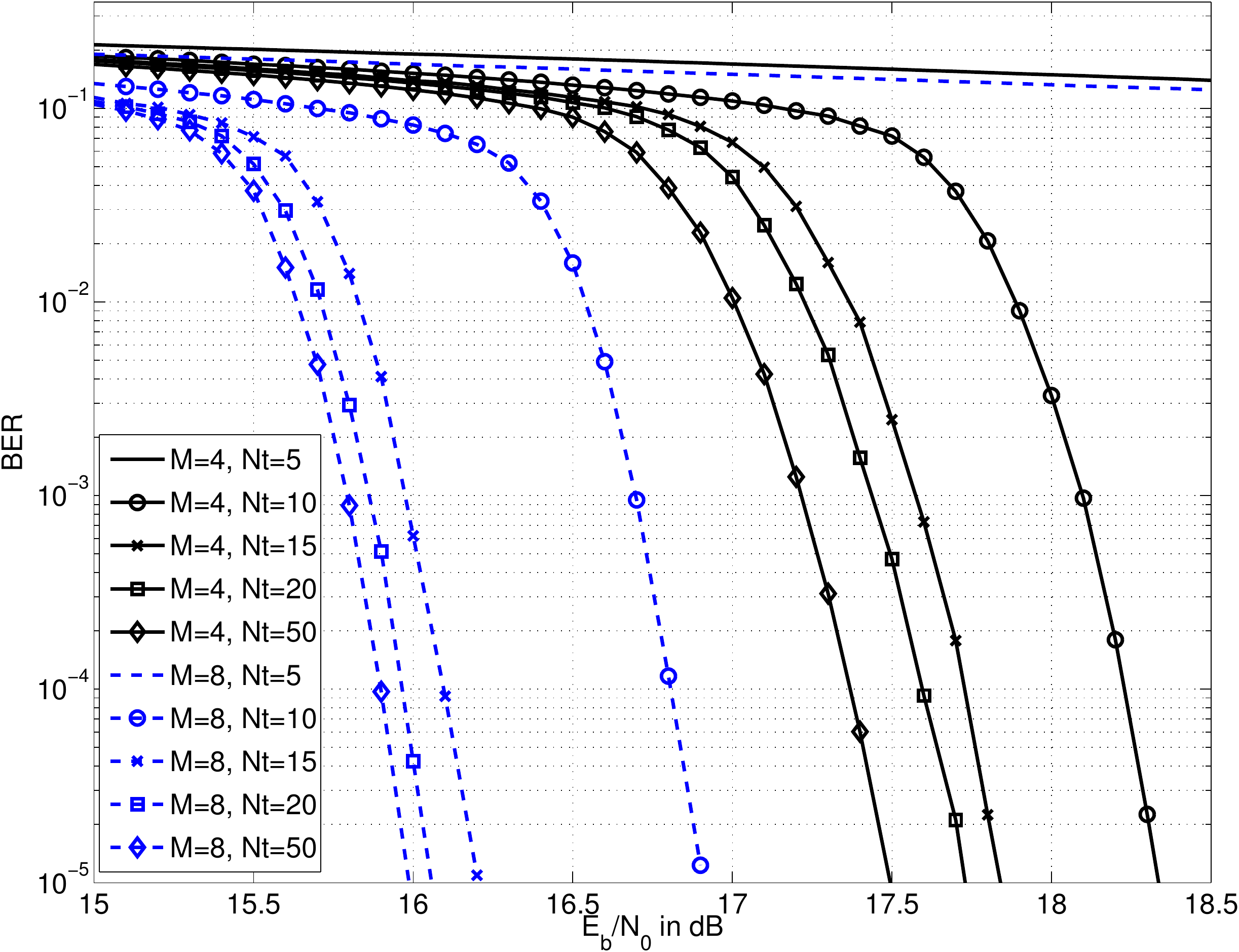}
   \caption{LDPC coded bit error rate performance at the end node in the two-way relay channel broadcast phase under Rayleigh
fading as a function of demodulator infinite series terms.  The LDPC code parameters are codeword length $L=16200$ and rate $r_S=1/2$.
The modulation orders considered are $M=\{4,8\}$
The number of demodulator infinite series terms considered are $N_t=\{5,15,25,50\}$.
All simulations use BICM decoding.
}
   \label{fig:coded_nt}
\vspace{-3mm}
 \end{figure}

Uncoded end node error rate performance as a function of modulation order and number of demodulator infinite series
terms is shown in Fig. \ref{fig:uncoded}.
For both modulation orders $M=2$ and $M=4$ and $N_t < 50$, a behavior is observed where detection fails completely 
after a particular SNR threshold is reached.
At $N_t=\{5,15,25\}$, the error threshold occurs at error rates $\approx 10^{-1}$, $\approx 10^{-3}$, and
$\approx 10^{-4}$ respectively.
For $N_t=50$, no threshold is observed for the error rates considered.
These results suggest that a minimum number of terms must be computed to operate at a particular error rate.

Channel coded error rate performance as a function of modulation order and number of infinite series terms
is shown in Fig. \ref{fig:coded_nt}.
In all cases, BICM with no decoder to demodulator feedback was used.
As in the uncoded case, performance is affected by the number of infinite series terms computed $N_t$, however,
an error threshold is only observed for the case $N_t=5$.
When channel coding is applied, the number of infinite series terms affects the location of the decoding waterfall region.
For modulation order $M=4$, the worst performing waterfall at $N_t=10$ is about $0.9$ dB worse than
the best performing waterfall at $N_t=50$.
The same difference is observed for modulation order $M=8$.
In the coded case, generally, fewer infinite series terms are required for successful decoding than in the uncoded case,
suggesting a tradeoff between demodulation and decoding complexity.

\begin{figure}[!t]
  \centering
  \includegraphics[width=9cm]{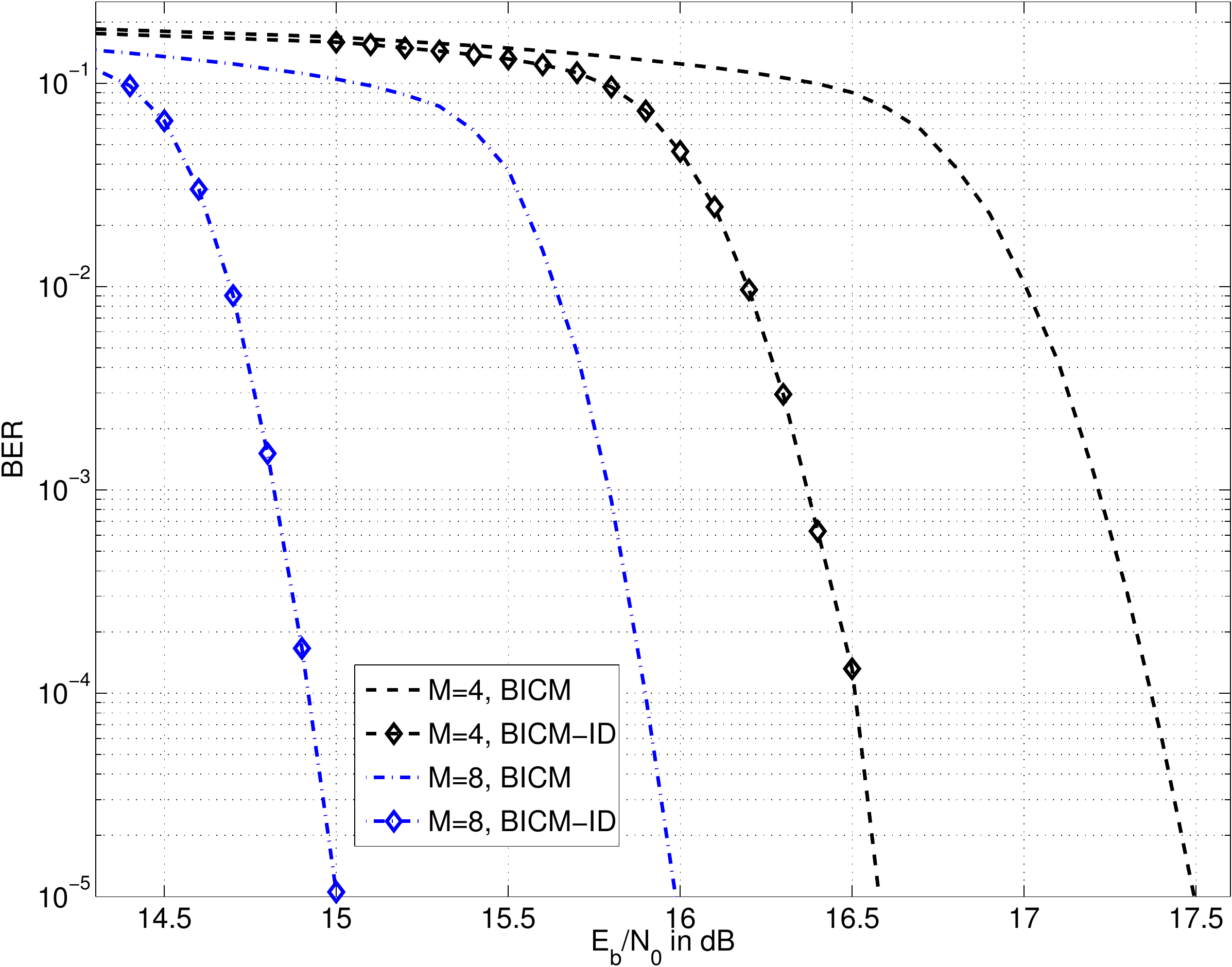}
  \caption{LDPC coded bit error rate performance at the end node in the two-way relay channel broadcast phase under Rayleigh
fading as a function of decoder feedback (BICM vs BICM-ID).  The LDPC code parameters are codeword length $L=16200$ and rate $r_S=1/2$.
The modulation orders considered are $M=\{4,8\}$.
All codes are simulated using $N_t = 50$ infinite series terms at the demodulator.}
  \label{fig:fig_coded_maxperf}
\vspace{-3mm}
\end{figure}

Channel coded error rate performance as a function of modulation order and decoder feedback is shown in
Fig. \ref{fig:fig_coded_maxperf}.
All codes are simulated using $N_t = 50$ infinite series terms at the demodulator.
The purpose of this experiment is to investigate the performance benefit yielded by information feedback
from decoder to demodulator, and the absolute performance difference between modulation orders $M=4$ and $M=8$.
For modulation order $M=4$, the BICM-ID exhibits a performance gain of $0.9$ dB over BICM.
For $M=8$, BICM-ID exhibits a gain of $1$ dB.
BICM for $M=8$ outperforms BICM for $M=4$ by approximately $1.5$ dB.

\balance

\section{Conclusion}

 This work developed a noncoherent soft output FSK demodulator the end nodes in the analog network-coded two-way
 relay channel under Rayleigh fading.
 The demodulator supports power of two modulation orders and iteration with the channel decoder.
 The demodulator formulation contains an infinite series which must be truncated for practical
 receiver implementation.
 It is demonstrated the bit error rate performance is sensitive to the infinite series truncation length.
 An exact characterization of the convergence of the demodulator as well as a closed form expression
 are left as future work.

\bibliographystyle{IEEEtran}
\bibliography{bibliography}

\end{document}